\documentclass[twocolumn]{revtex4-2}
\usepackage{amsmath}
\usepackage{bm}
\usepackage{graphicx}
\usepackage{float}
\usepackage{subcaption}
\usepackage{xcolor} 
\usepackage[colorlinks=true, linkcolor=blue, citecolor=blue, urlcolor=blue]{hyperref} 

\begin{document}

\title{Quantum Classical Correspondence Using Coherent State Measurements and Husimi Q Probability Distributions}

\author{Youheng Zheng}
\affiliation{White Station High School, Memphis, TN, USA}

\date{August 24th, 2025}

\begin{abstract}

We propose and simulate a protocol to evolve a quantum particle forward in time such that its trajectory closely matches that of the particle's Newtonian counterpart. Using short bursts of Schrödinger time-evolution interleaved with positive operator-valued measurements (POVMs) in the coherent basis, we demonstrate quantum-classical convergence for durations far beyond Schrödinger time-evolution alone. We examine the impact of the time between measurements $\Delta t$ and the reduced Planck's constant $\hbar$ on divergence time. Results indicate that for appropriate values of $\Delta t$, smaller values of $\hbar$ lead to longer divergence times. This method suggests a elegant, intuitive bridge to recover classical motion from quantum postulates. 
\end{abstract}

\maketitle

\section{Introduction}

Understanding the classical limit of quantum mechanics remains a fundamental challenge in modern physics. Classical systems are deterministic, whereas quantum systems evolve probabilistically, leading to conflicting models of motion. Heisenberg's uncertainty principle, which forbids simultaneous precision in a particle’s position and momentum, makes quantum measurements inherently probabilistic \cite{Heisenberg}. In contrast, a classical particle under the same initial conditions would have an exact and predictable trajectory in phase space. Under simple Schrödinger evolution, the path of a quantum particle rapidly diverges from its classical counterpart. However, we expect classical mechanics to emerge from quantum theory in the appropriate limit, a principle known as quantum–classical correspondence.

To further understand this, various ideas have been proposed over the years. Notable examples include Ehrenfest's theorem and the decoherence theory \cite{Decoherence paper}. However, Ehrenfest's theorem fails quickly, while the decoherence theory relies on heavy, elaborate math. Our method is a simple and lean way to recover quantum-classical correspondence for periods of time longer than Ehrenfest's theorem without the cumbersome math of the decoherence theory. 

Our method combines short bursts of time-evolution with quantum measurements over positive operator-valued measures (POVMs) \cite{POVM lecture}. The bursts of time evolution allow a particle to move forward in phase space, yet prohibits too much spread. After each timestep, we instantaneously generate a Husimi Q probability distribution: a smooth, nonegative representation of a quantum state in phase space. We then sample a random value of position and momentum from it, accounting for quantum's probabilistic nature. Lastly, we collapse the state vector into a coherent state centered at this point. By repeating this procedure, we achieve long periods of convergence between quantum and classical particles across a variety of one-dimensional potentials. 

\section{Methods}

The core of our method is as follows:
\begin{enumerate}
\item Initiate a quantum and classical particle with the same initial position, momentum, and potential. The quantum particle is initialized in a coherent state.
\item Time-evolve the classical particles under Newtonian mechanics
\item Schrödinger time-evolve the quantum particle for short bursts of time $\Delta t$
\item Between each burst, generate a Husimi Q probability distribution for the state
\item Randomly sample a point from the distribution
\item Collapse the state vector into a coherent state centered at this sampled point
\item Repeat until the divergence threshold is exceeded
\end{enumerate}

A guiding principle of our method is its straightforwardness. We employ direct Schrödinger time-evolution to simulate the quantum particle. Random sampling from the Husimi Q distribution (figure \ref{fig: husimi q}) captures the inherent randomness of quantum mechanics, while collapsing the state vector into a coherent state minimizes uncertainty.
\section{Simulation}
\subsection{Physics Setup}
The simulation initializes by defining the physical parameters of both the classical and quantum systems, including particle mass, initial position, initial momentum, and characteristic frequencies. Various potential energy landscapes are implemented, encompassing common test cases such as harmonic, free, linear, quartic, Gaussian well, and double-well potentials. Each potential is represented mathematically through a potential function \(V(x)\) and its corresponding force \(F(x) = -\frac{dV}{dx}\). Quantum states are prepared as coherent states, which are Gaussian wavepackets with minimal uncertainty in position and momentum.\cite{Anatoli}

\subsection{Numerical Techniques}
\begin{figure}[H]
    \begin{minipage}[c]{0.5\linewidth}
        \includegraphics[width=\linewidth]{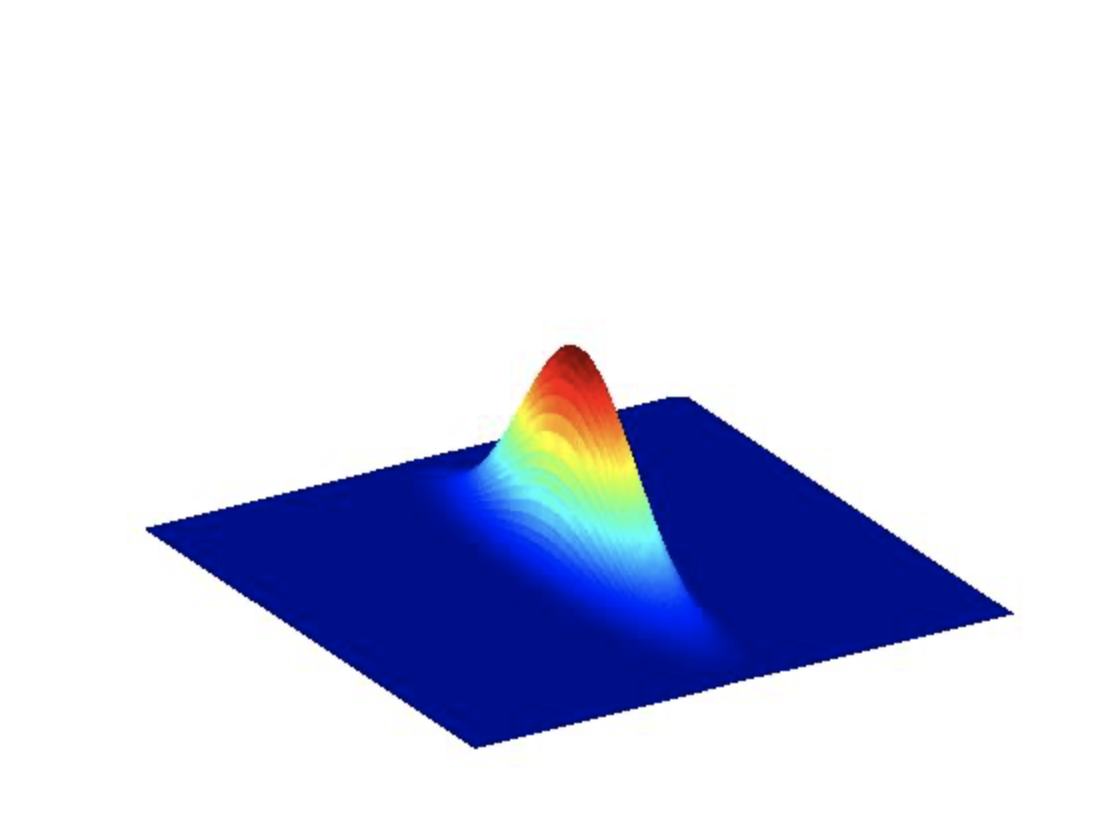}
        \caption{Husimi Q \cite{Wikipedia}}
        \label{fig: husimi q}
    \end{minipage}
    \hfill
    \begin{minipage}[c]{0.4\linewidth}
        \includegraphics[width=\linewidth]{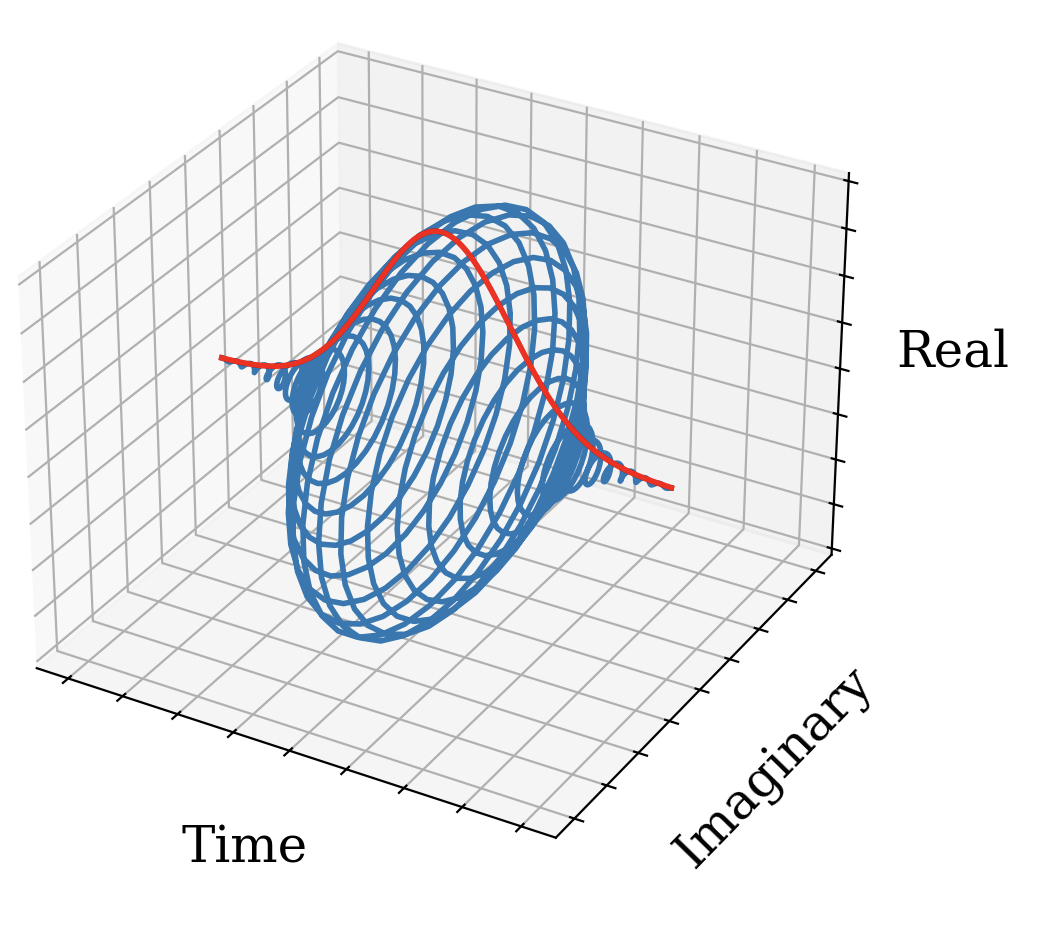}
        \caption{Coherent State}
        \label{fig: coherent state}
    \end{minipage}
\end{figure}

Quantum dynamics are propagated using the split-operator method, which separates the time evolution into potential and kinetic contributions. This approach applies half a potential step in position space, performs a Fourier transform to momentum space to apply kinetic evolution, and then returns to position space to complete the remaining half-step. The method preserves unitarity, achieves third-order accuracy in the timestep, and reduces computational complexity from \(O(N^3)\) to \(O(N \log N)\) by leveraging fast Fourier transforms. Classical trajectories are integrated with a fourth-order Runge-Kutta scheme, providing high-accuracy solutions to Newton's equations of motion. Ensemble quantum simulations are organized with the timestep forming the outer loop and the ensemble member index forming the inner loop, ensuring consistent comparisons at each evolution step. Partial measurements of position and momentum are taken through positive operator-valued measurements (POVMs) \cite{POVM lecture}by sampling the Husimi \(Q\) distribution [\ref{fig: husimi q}] at each timestep and collapsing the wavefunction into a coherent state [\ref{fig: coherent state}] centered at the sampled point.

\subsection{Data Handling and Divergence Metric}
Simulation data is recorded using structured outputs to capture both physical and numerical information. Trajectory information is stored as tuples of time, position, and momentum. The phase-space distribution is represented using the Husimi Q function, which is recomputed at each timestep to allow sampling and to ensure physical interpretability. A divergence metric quantifies the ensemble-averaged deviation of quantum trajectories from the classical reference in phase space: the root-mean-square (RMS) deviation $D(t)$ is calculated as
\[
D(t) = \sqrt{ 
\frac{1}{N} \sum_{i=1}^{N} \Big[ 
\big( x_{\mathrm{c}}(t) - x_i(t) \big)^2 
+ \big( p_{\mathrm{c}}(t) - p_i(t) \big)^2 
\Big] 
}, 
\]
where $N$ is the number of quantum runs being compared, $x_c(t)$ and $p_c(t)$ are the position and momentum of the classical particle at time $t$, and $x_i(t)$ and $p_i(t)$ are the position and momentum of the $i$th quantum particle, as measured at time $t$. 

This captures both the spread of the ensemble and deviations induced by quantum effects, including backaction from repeated Husimi-Q sampling. The simulation terminates if this RMS deviation exceeds a predefined threshold, providing a systematic measure of when quantum-classical correspondence weakens. Wavefunction grids are dynamically adjusted using moving windows to follow the particle, with linear interpolation preserving normalization, streamlining computational efficiency. Husimi \(Q\) windows are reconstructed at each step, avoiding interpolation artifacts.

\subsection{Computational Implementation}
The simulation is implemented in Python and executed on a high-performance computing cluster. Large-scale matrix operations, including Fourier transforms and Hilbert-space manipulations, require significant computational resources. To explore the dependence of divergence on both timestep and reduced Planck's constant, parameter sweeps are performed using parallel computation. Each core handles a unique combination of timestep and Planck’s constant, producing data that are subsequently compiled into heatmap and phase-space plots. These heatmaps provide a visual representation of divergence times across the parameter space, with each pixel corresponding to a single core’s simulation output. This architecture allows efficient batch processing and facilitates exploration of the system’s physical behavior.

\subsection{Parameters}
In this simulation, we discretized $\Delta{t}$ and $\hbar$ into 25 values each to produce a 625-pixel heatmap using the following parameters: 

\begin{table}[h]
\centering
\begin{tabular}{ll}
\hline
Parameter & Value \\
\hline
Time step, $\Delta t$ & 0.01 -- 0.3 s \\
Planck constant, $\hbar$ & $3.0\times 10^{-6}$ -- $1.0\times 10^{-2}$ \\
Mass, $m$ & 1 \\
Oscillator frequency, $\omega$ & 1 \\
Classical timestep & 0.01 s \\
Initial position, $x_0$ & 0 \\
Initial momentum, $p_0$ & 1 \\
Potential function $V(x)$ & $\frac{5}{2}x^2$ \\
Quantum ensemble size & 25 \\
Momentum prefactor & 8 \\
Uncertainty prefactor & 15 \\
Divergence threshold & 0.05 \\
Husimi Resolution & 50\\
\hline
\end{tabular}
\end{table}

We adopted dimensionless natural units $m = \omega = 1$ to simplify the system. Ensemble size is the number of quantum runs we simulated for each pair of $\Delta{t}$ and $\hbar$; a larger ensemble size reduces stochasticity. We then averaged divergence time across all runs. The momentum prefactor controls how much space the computational window allocates for particle motion between measurements, while the uncertainty prefactor ensures the window is wide enough to capture quantum spreading. Our algorithm selects whichever requirement is more stringent to prevent wavefunction truncation while maintaining computational efficiency.

\section{Delineating Regimes}

The time evolution of the probability distribution $P(x, p, t)$ for a classical particle in phase space is governed by the Poisson bracket with the Hamiltonian. Specifically, the rate of change of $P$ is given by
\[
\frac{\mathrm{d}P}{\mathrm{d}t} = \{ H, P \},
\]
\\
where $H(x, p)$ is the classical Hamiltonian of the system, typically representing the total energy, and $\{ H, P \}$ denotes the Poisson bracket, defined as
\[
\{ H, P \} = \frac{\partial H}{\partial x} \frac{\partial P}{\partial p} - \frac{\partial H}{\partial p} \frac{\partial P}{\partial x}.
\]

The Poisson bracket encodes how one observable evolves under the dynamics generated by another. In this case, it represents the evolution of the probability density $P(x, p, t)$ under the Hamiltonian $H$. It serves as the classical analog of the quantum commutator.

The Wigner function $W(x, p, t)$ is a quasi-probability distribution defined over phase space, constructed to represent quantum states in a way that parallels classical probability densities. Unlike the classical probability density $P(x, p, t)$, the Wigner function can take on negative values, reflecting the non-classical nature of quantum interference.

In the quantum analogue of classical phase space dynamics, the Wigner function $W(x, p, t)$ plays the role of a quasi-probability distribution. The time evolution of $W$ is governed by the Moyal bracket with the Hamiltonian, analogous to the classical Poisson bracket. Specifically,
\[
\frac{\mathrm{d}W}{\mathrm{d}t} = \{ H, W \}_{\text{M}},
\]
where $\{ H, W \}_{\text{M}}$ denotes the Moyal bracket, defined as
\[
\{ A, B \}_{\text{M}} = \frac{2}{\hbar} A \sin\left( \frac{\hbar}{2} \left( \overleftarrow{\partial}_x \overrightarrow{\partial}_p - \overleftarrow{\partial}_p \overrightarrow{\partial}_x \right) \right) B.
\]

Here, the function $H(x, p)$ represents the Weyl transform of the quantum Hamiltonian operator $\hat{H}$. This transformation is necessary to express operator dynamics in phase space, ensuring the correct symmetrization of non-commuting operator terms. In cases where $\hat{H}$ contains mixed products like $\hat{x} \hat{p}$, the Weyl transform yields a properly symmetrized function in $x$ and $p$, allowing for accurate quantum corrections in the Moyal evolution.

The arrows indicate the direction of differentiation, and the sine operator should be understood as a formal power series expansion in derivatives.

We begin with the Moyal bracket form of the Wigner equation,  
\begin{equation*}
\frac{\mathrm{d}W}{\mathrm{d}t} 
= \frac{2}{\hbar} H \sin\!\left( \tfrac{\hbar}{2} 
\big( \overleftarrow{\partial}_x \overrightarrow{\partial}_p 
- \overleftarrow{\partial}_p \overrightarrow{\partial}_x \big) \right) W.
\end{equation*}

Expanding the sine to lowest nontrivial order gives
\begin{equation*}
\frac{\mathrm{d}W}{\mathrm{d}t} =
\frac{\partial H}{\partial x} \frac{\partial W}{\partial p}
- \frac{\partial H}{\partial p} \frac{\partial W}{\partial x}
- \frac{\hbar^2}{24} \, \mathcal{D}[H,W] 
+ \mathcal{O}(\hbar^4),
\end{equation*}
where the operator $\mathcal{D}[H,W]$ collects cubic derivative terms in $x$ and $p$.
The first two terms reproduce the classical Liouville equation, while the 
$\hbar^2$ correction quantifies the leading quantum deviation. 
Substituting first, second, and third order differentials of $H$ and $W$ leads us to the asymptotal inequality: 
\begin{align*}
&\boxed{\frac{(3\hbar m \omega\delta p - 4\delta p^3) V'''(x)}
{6m^2\omega^2\big(m\omega^2 \delta x - \delta pV'(x)\big)} \;\ll\; 1}
\end{align*}
The full derivation is provided in the Appendix.
Another important regime to consider is the uncertainty-dominated limit. 
Here, overly frequent measurement introduces additional randomness in the 
wavepacket’s position, effectively broadening its displacement beyond the 
classical drift. Over a single timestep $\Delta t$, the classical center 
moves by
\[
\Delta x_{\text{classical}} \sim \frac{p}{m}\Delta t,
\]
while the measurement process induces a stochastic shift of order
\[
\Delta x_{\text{measurement}} \sim \sqrt{\frac{\hbar}{2 m \omega}}.
\]
Since the actual displacement is the sum of these two contributions, 
consistency requires that the measurement-induced jump remain small 
compared to the classical drift,
\[
\sqrt{\frac{\hbar}{2 m \omega}} \;\ll\; \frac{p}{m}\,\Delta t.
\]

\noindent Rearranging yields
\[
\boxed{\frac{\hbar m}{2 \omega p^2 \Delta t^2} \;\ll\; 1},
\]

This condition ensures that quantum fluctuations from measurement remain 
subdominant to the deterministic classical motion of the wavepacket center.

For parameters that satisfy both equations, we expect to observe classical behavior. 
\section{POVM Framework}

The following construction rigorously justifies using the Husimi Q function as a probability distribution over phase space. Coherent states, which form an overcomplete, non-orthogonal basis, do not correspond to standard projective measurements, so directly measuring them would violate the postulates of quantum mechanics. To resolve this, we embed the original Hilbert space $\mathcal{H}_a$ into a larger space $\mathcal{H} = \mathcal{H}_a \otimes \mathcal{H}_b$ via an isometry. This allows the overcomplete set of coherent states to be extended to a complete orthonormal basis in the enlarged space, making standard projective measurements possible. The proof that follows shows explicitly how this embedding reproduces a valid POVM on $\mathcal{H}_a$ and leads, in the continuous limit, to the Husimi Q distribution. \\

Define $\mathcal{H}_a$ as the Hilbert space spanned by the orthonormal set of position eigenstates
\[
\{\, |x_1\rangle, |x_2\rangle, \dots, |x_k\rangle \,\},
\]
or equivalently the set of momentum eigenstates
$$\{ \ |p_1\rangle, |p_2\rangle, \dots, |p_k\rangle \}$$

To construct a probability distribution over both position and momentum (phase space), we introduce coherent states, which are labeled by $(x_0, p_0)$ and represent minimally uncertain, “classical-like” wavepackets. Coherent states are specific linear combinations of these basis states that minimize the uncertainty in both position and momentum. Physically, the state $|x_0, p_0\rangle$ is localized near position $x_0$ and momentum $p_0$, representing a fuzzy but classical-like point in phase space:
\[
\langle x | x_0, p_0 \rangle = \left( \frac{m\omega}{\pi \hbar} \right)^{\!1/4}
\exp\left[ -\frac{m\omega}{2\hbar} (x - x_0)^2 + i \frac{p_0 x}{\hbar} \right].
\]

For mathematical convenience, we discretize phase space into $k$ positions and $k$ momenta, producing a finite set of $k^2$ coherent states:
\[
\{\, |x_1,p_1\rangle, |x_1, p_2\rangle, \dots, |x_1, p_k\rangle, |x_2, p_1\rangle, \dots |x_k, p_k\rangle \,\}.
\]

To make the overcomplete set of coherent states compatible with standard projective measurements, we embed $\mathcal{H}_a$ into a larger Hilbert space via an isometry. Define $\mathcal{H}_b$ as an expanded Hilbert space of dimension $k^2$, spanned by the orthonormal set
\[
\{\, |E_1\rangle, |E_2\rangle, \dots, |E_{k^2}\rangle \,\}.
\]
We construct an isometry
\[
V : \mathcal{H}_a \to \mathcal{H} := \mathcal{H}_a \otimes \mathcal{H}_b,
\]
which embeds $\mathcal{H}_a$ into the larger space $\mathcal{H}$, allowing the overcomplete basis of coherent states to be extended to a complete orthonormal basis. The isometry and its adjoint are
\[
V = \sum_{i=1}^{k^2} \sqrt{c_i} \, |E_i\rangle \langle \phi_i |, \quad
V^\dagger = \sum_{j=1}^{k^2} \sqrt{c_j} \, |\phi_j\rangle \langle E_j |,
\]
where $c_i > 0$ are weights chosen to ensure proper normalization of the resulting POVM. Indeed, the POVM normalization follows from
\[
V^\dagger V = \sum_{i,j=1}^{k^2} \sqrt{c_i c_j} \, |\phi_j\rangle \langle E_j | E_i \rangle \langle \phi_i |
= \sum_{i=1}^{k^2} c_i \, |\phi_i\rangle \langle \phi_i | = I.
\]
Given $|\psi\rangle \in \mathcal{H}_a$, define $\rho = V |\psi\rangle \langle \psi| V^\dagger$. Tracing out $\mathcal{H}_b$ gives the reduced state
\[
\rho_A = \mathrm{Tr}_B(\rho) = \sum_{i=1}^{k^2} c_i \, |\phi_i\rangle \langle \phi_i | \, |\langle \phi_i | \psi \rangle|^2.
\]
The measurement statistics become
\[
P(i) = \mathrm{Tr}(\rho_A \, |\phi_i\rangle \langle \phi_i|) = c_i |\langle \phi_i | \psi \rangle|^2.
\]
Taking the continuous limit of infinitely fine discretization, the discrete POVM reproduces the standard Husimi $Q$ distribution over phase space:
\[
Q(x, p) = \frac{1}{\pi} \left| \langle x, p | \psi \rangle \right|^2.
\]
The prefactor $1/\pi$ arises because in the limit of infinitely dense phase-space coverage, the weights $c_i$ become $dx\,dp / \pi$, ensuring the normalization
\[
\int dx \, dp \, Q(x,p) = 1.
\]

\section{Results and Discussion}

In this section, we show simulation results for different values of $(\hbar, \Delta{t})$ and how they impact divergence time. 
\begin{figure}[H]
    \raggedright
    \includegraphics[width=0.85\columnwidth]{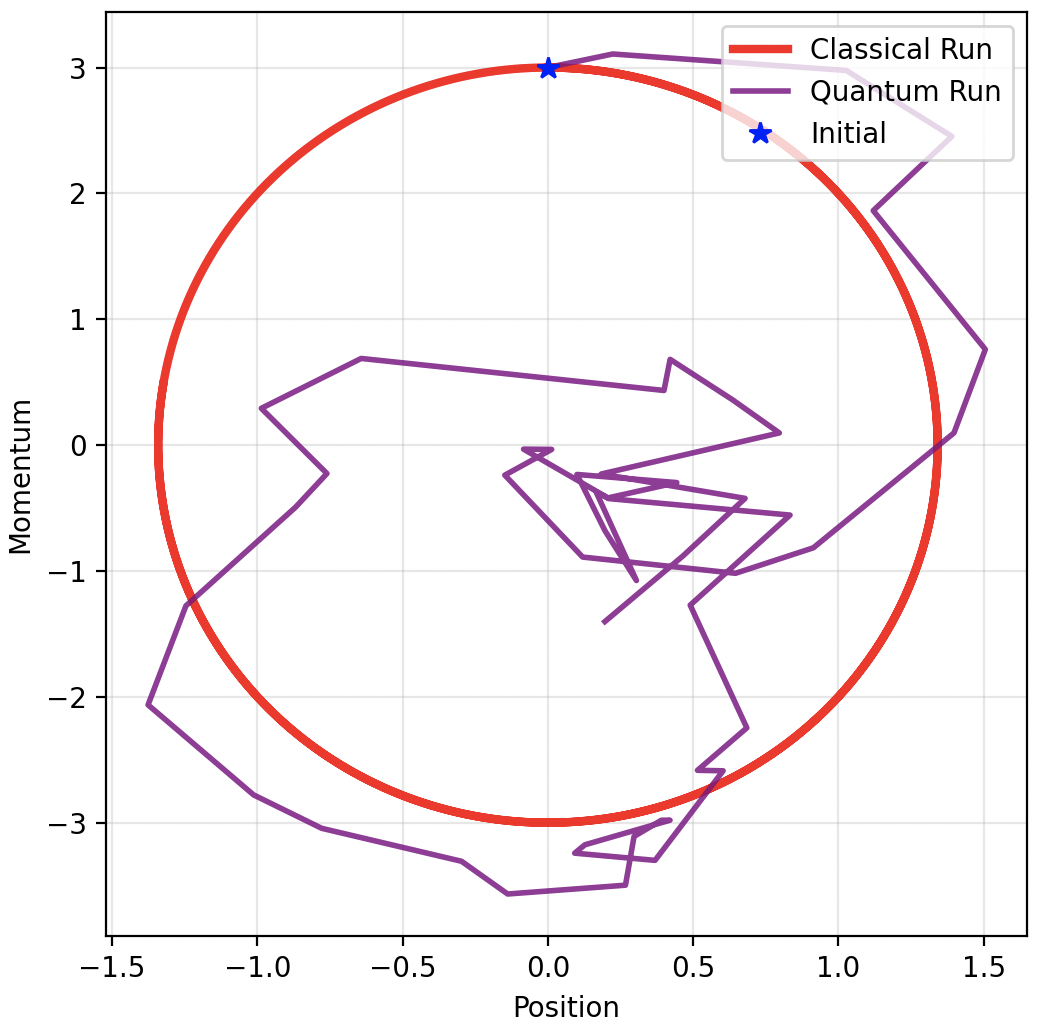}
    \caption{$\hbar=0.1$, $\Delta t=0.1$}
    \label{fig:bad_phase}

    \includegraphics[width=0.85\columnwidth]{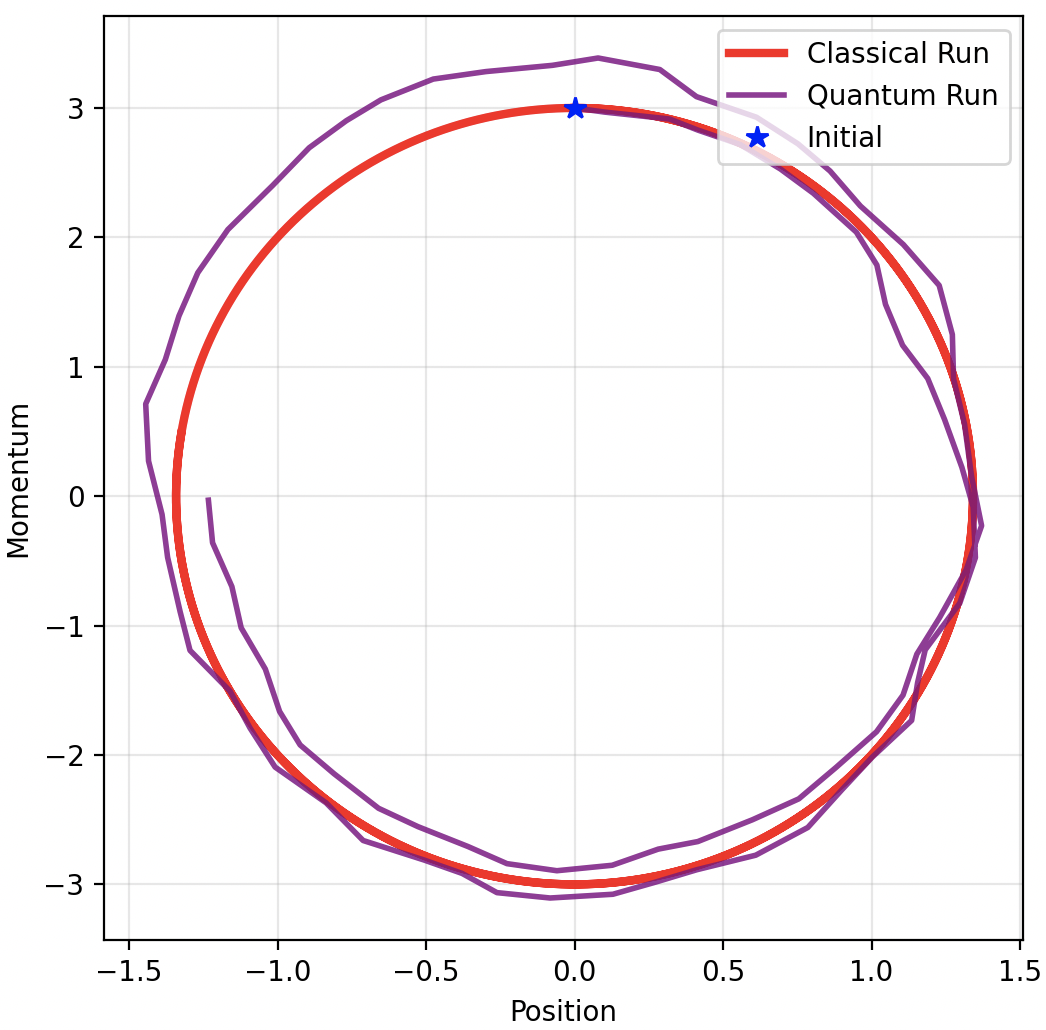}
    \caption{$\hbar=0.001$, $\Delta t=0.055$}
    \label{fig:mid_phase}

    \includegraphics[width=0.85\columnwidth]{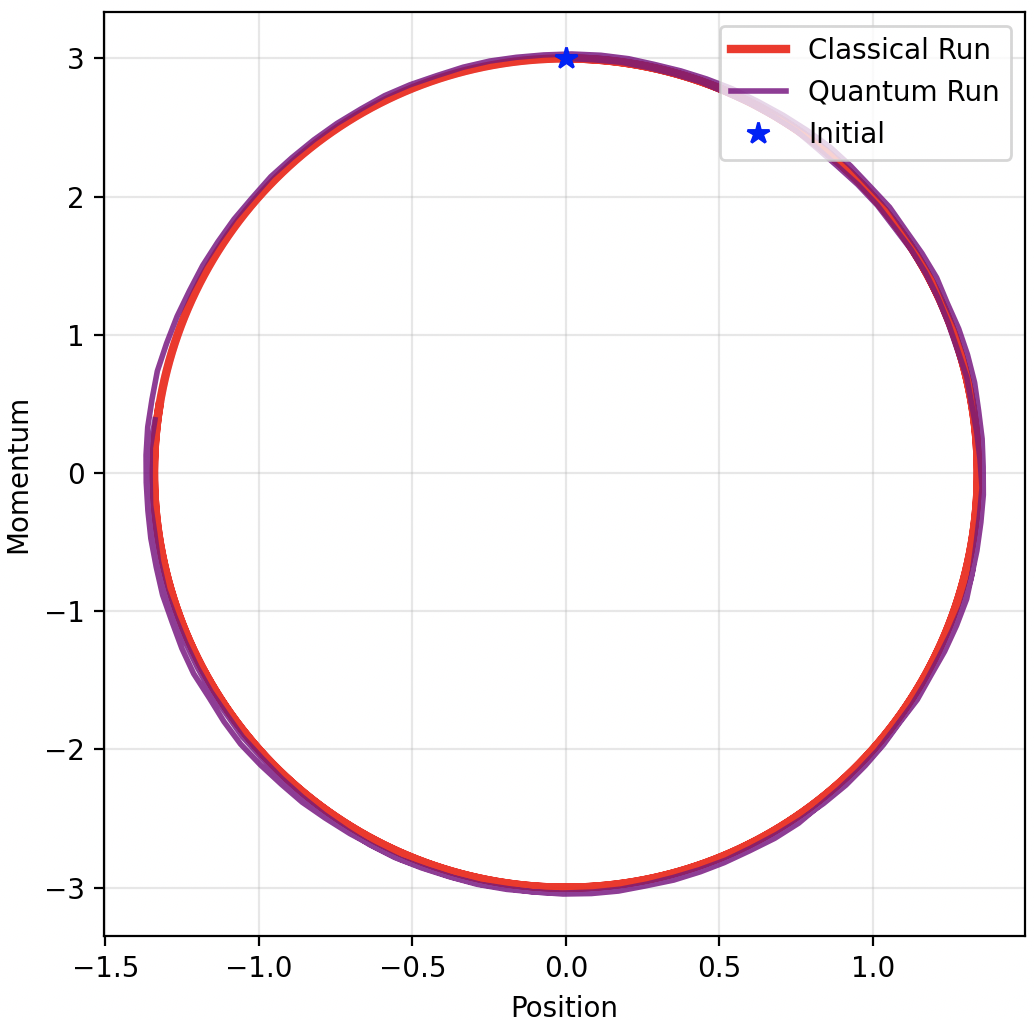}
    \caption{$\hbar=0.00001$, $\Delta t=0.03$}
    \label{fig:good_phase}
    
\end{figure}

\begin{figure}[H]
    \includegraphics[width=\columnwidth]{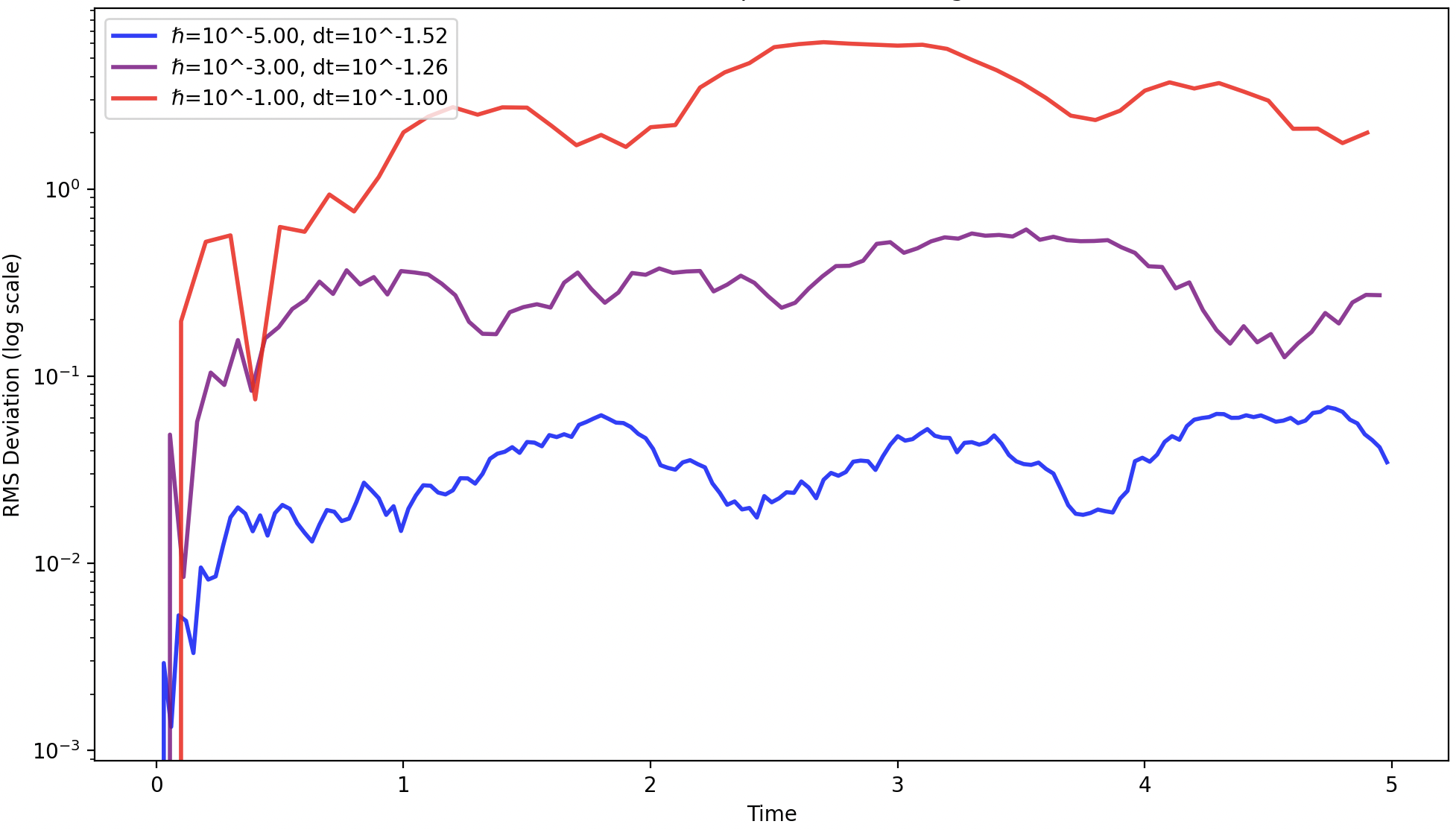}
    \caption{Root-mean-square (RMS) deviation over time}
    \label{fig:rms}
\end{figure}

\begin{figure}[H]
    \includegraphics[width=\columnwidth]{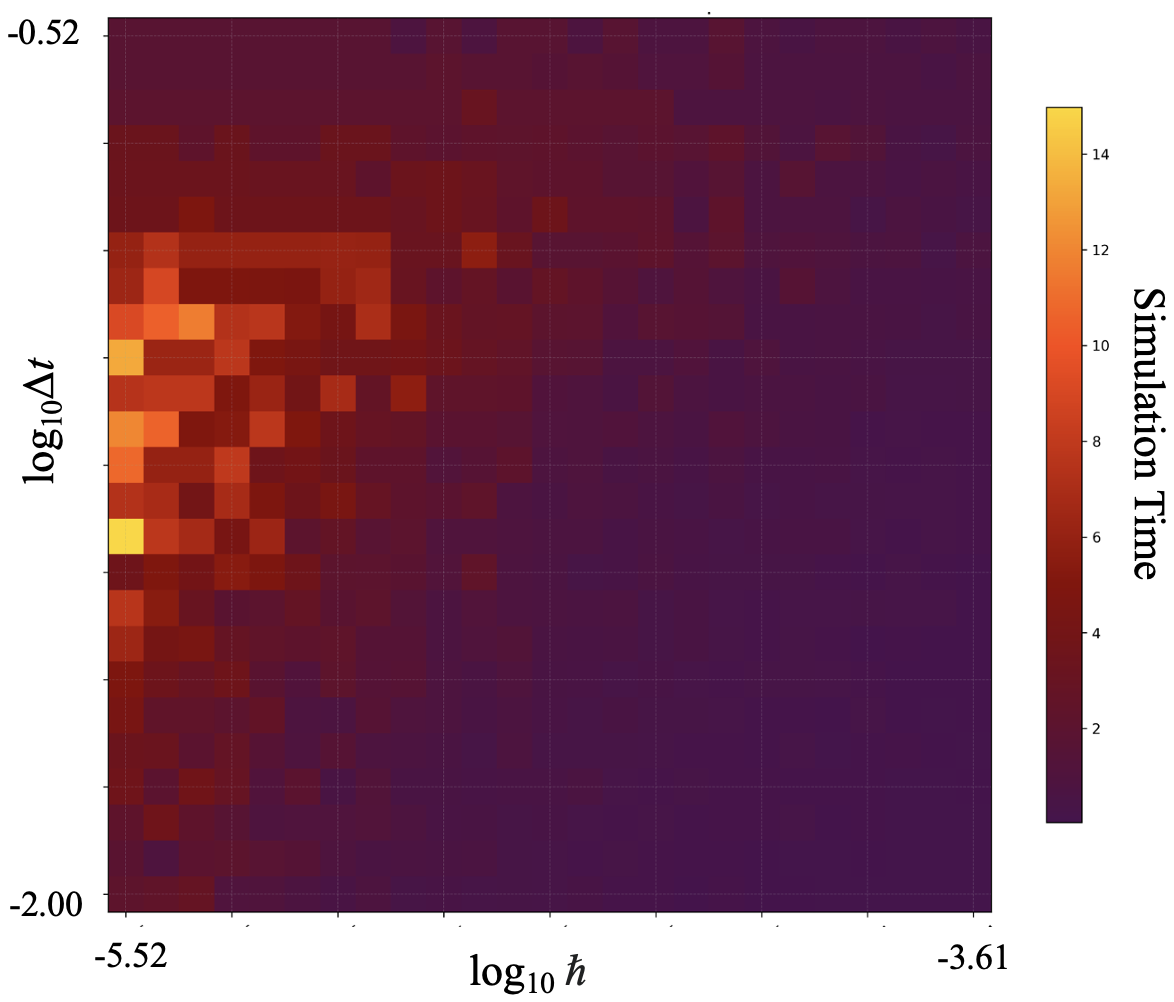}
    \caption{Divergence time heatmap}
    \label{fig:final_heatmap}
\end{figure}

\begin{figure}[H]
    \includegraphics[width=0.84\columnwidth]{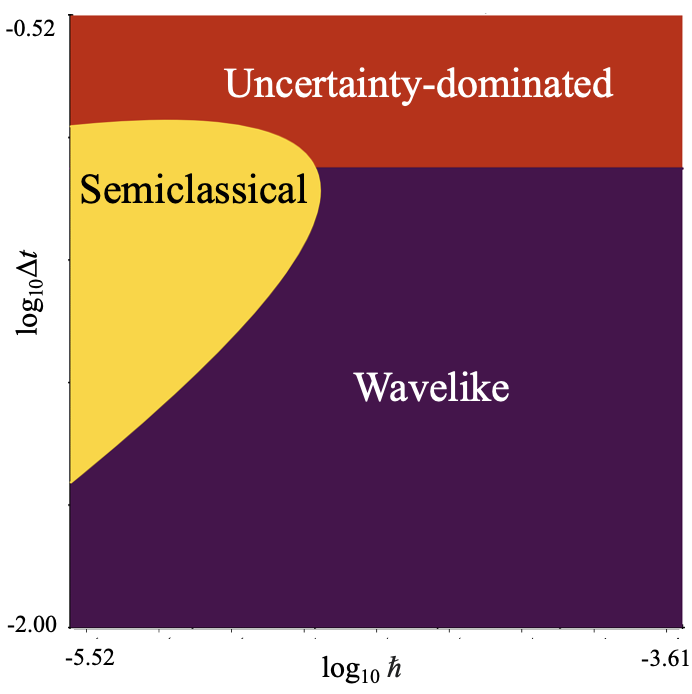}
    \caption{Heatmap regimes}
    \label{fig:regime_outline}
\end{figure}

Figures \ref{fig:bad_phase} through \ref{fig:good_phase} illustrate the phase space trajectories of a classical particle in a harmonic potential, and a quantum particle initialized to a coherent state centered at the classical particle's initial position and momentum.  The reported values of quantum particle's position and momentum are the sampled values from the coherent measurements. The purple line represents the quantum particle's trajectory while the red line represents the classical particle's trajectory. 

We observed that as $\hbar \to 0$, quantum and classical trajectories show higher levels of convergence and closer alignment in phase space.

For $\hbar=0.1, \Delta{t}=0.1$, the quantum particle immediately diverges from the classical particle, as evidenced by their completely different trajectories. For $\hbar=0.001, \Delta{t}=0.055$, the quantum particle diverges a smaller but considerable amount from the classical particle, yet stayed relatively close; their trajectories in phase space take on nearly the same shape. For $\hbar=0.00001, \Delta{t}=0.03$, the quantum particle exhibits high levels of correspondence, shown by the almost complete overlap of the red and purple lines. 

Figure \ref{fig:rms} illustrates the impact of $\hbar$ and $\Delta{t}$ on RMS deviation over time. For $\hbar=0.1, \Delta{t}=0.1$, the RMS deviation has magnitude $10^{0}$, with a maximum of around 6. For $\hbar=0.001, \Delta{t}=0.055$, the RMS deviation has magnitude $10^{-1}$, with a maximum of around 0.6. For $\hbar=0.00001, \Delta{t}=0.03$, the RMS deviation has magnitude $10^{-2}$, with a maximum of around 0.07. Furthermore, we see that as $\hbar$ decreases by two magnitudes, the scale of RMS deviation decreases by one magnitude. 

All three RMS deviation plots show semi-periodic increases and decreases. These correspond with the slightly more divergent spots on the "ends" the phase space graph, where momentum equals 0. This physical phenomenon occurs at the two ends of the harmonic potential, where the particle momentarily achieves zero momentum while reversing direction. 

Figures \ref{fig:final_heatmap} and \ref{fig:regime_outline} show the impact of $\hbar$ and $\Delta{t}$ on divergence time. Figure \ref{fig:final_heatmap} contains a 25-by-25 grid of pixels, each one representing a combination of parameters $(\hbar,\Delta{t})$. The color of the pixels represent the time until the quantum and classical run RMS divergence exceeded 0.05. The second outlines the regime boundaries. Parameters are spaced logarithmically to study multiple orders of magnitude effectively. Due to computational restrictions, we were only able to lower $\hbar$ to about $10^{-5.5}$. 

As $\hbar$ decreases, divergence time increases significantly. In every row, pixels became lighter in shade from right to left. However, this is not the case with $\Delta{t}$. Given the same value of $\hbar$, as $\Delta{t}$ decreases, divergence time first increases, then decreases. For $\hbar=3.00 \times 10^{-6}$, $\Delta{t=0.30}$ yields a divergence time of approximately 2 seconds, $\Delta{t=0.041}$ yields a divergence time of approximately 15 seconds, and $\Delta{t=0.01}$ yields a divergence time of approximately 3 seconds. In other columns on the left, we observed the same trend. 

We identified three regions, each representing a different regime, outlined by Figure \ref{fig:regime_outline}.

\subsection*{Uncertainty-dominated regime}

In the uncertainty-dominated regime, the time between measurements is very short. As a result, the Gaussian only moves a tiny distance in phase space, resulting in significant overlap with its previous position, illustrated by the image below. Then, when we sample, we are likely to select a point $(x,p)$ in the overlapping region. Hence, the particle essentially becomes "stuck", fluctuating locally instead of undergoing significant motion. This is mathematically explained by the inequality from earlier: 
\[
\frac{\hbar m}{2 \omega p^2 \Delta t^2} \;\ll\; 1
\]
As $\Delta t$ becomes too small, the left-hand side becomes too large, thus violating the inequality.

\begin{figure}[H]
    \includegraphics[width=1\columnwidth]{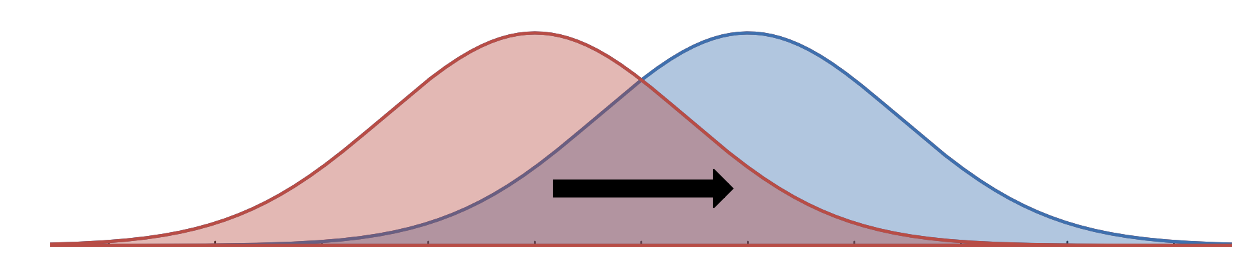}
    \caption{2D illustration of the uncertainty-dominated regime}
    \label{fig:uncertainty_regime}
\end{figure}

\subsection*{Wavelike regime}
In the wavelike regime, the time between measurements is too long. As a result, the Gaussian spreads significantly and exhibits wavelike behavior. Interference patterns also emerge, highlighted by the blue wavepacket in the diagram below. Infrequent measurement allows a quantum particle to progressively diverge from its classical counterpart. This is mathematically explained by the inequality from earlier: 
\[
\frac{(3\hbar m \omega\delta p - 4\delta p^3) V'''(x)}
{6m^2\omega^2\big(m\omega^2 \delta x - \delta pV'(x)\big)} \;\ll\; 1
\]

\begin{figure}[H]
    \includegraphics[width=\columnwidth]{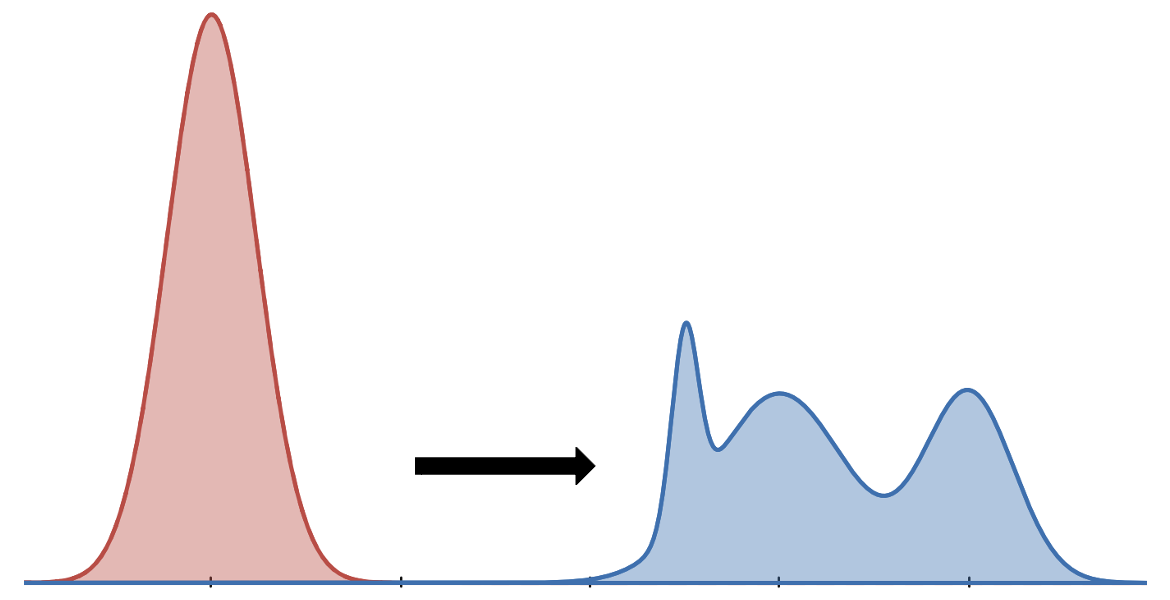}
    \caption{2D illustration of the wavelike regime}
    \label{fig:wavelike_regime}
\end{figure}

\begin{figure}[H]
    \includegraphics[width=\columnwidth]{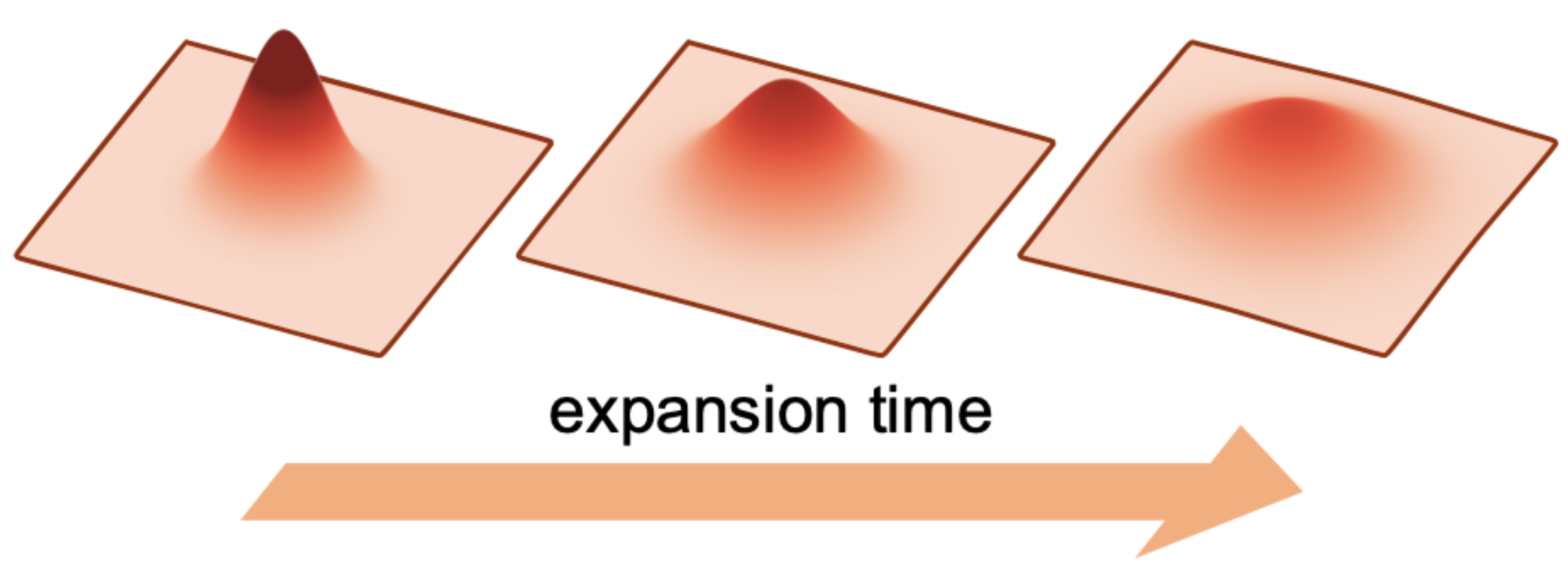}
    \caption{3D illustration of a Gaussian spreading \cite{Physics today}}
    \label{fig:wavelike_regime_3D}
\end{figure}

\subsection*{Semiclassical regime}
\begin{figure}[H]
    \includegraphics[width=\columnwidth]{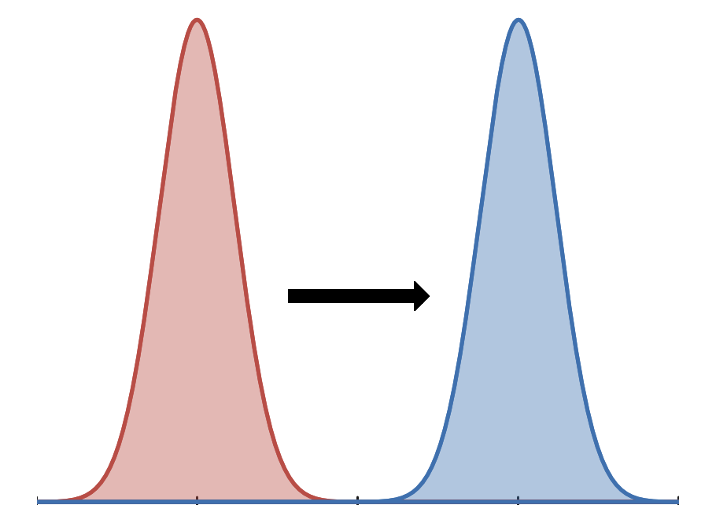}
    \caption{2D illustration of the semiclassical regime}
    \label{fig:semiclassical_regime}
\end{figure}
In a semiclassical regime, there is an ideal timestep between measurements. In this domain, we measure frequently enough to prevent wavelike spreading without inhibiting movement that respects the potential. This regime corresponds to well-tuned parameters for quantum-classical convergence, and hence had the longest divergence times.\\

The semiclassical regime must therefore satisfy both inequalities 
\[
\left\{
\begin{aligned}
&\frac{(3\hbar m \omega \, \delta p - 4 (\delta p)^3) V'''(x)}
{6 m^2 \omega^2 \big(m \omega^2 p \, \delta x - \delta p V'(x)\big)} \ll 1 \\
&\frac{\hbar m}{2 \omega p^2 \Delta t^2} \ll 1
\end{aligned}
\right.
\]
We see that as $\hbar \to 0$, both inequalities are satisfied. Our simulation data also supports this. Figure \ref{fig:rms} shows that decreasing $\hbar$ by 2 orders of magnitude decreases the RMS deviation by 1 order of magnitude; figure \ref{fig:final_heatmap} shows that as $\hbar$ decreases by approximately 2 magnitudes, from $10^{-3.61}$ to $10^{-5.52}$, divergence time increases by 1 magnitude, from $10^{0}$ to $10^{1}$. Therefore, we stipulate that in the context of macroscopic systems (where $\hbar$ is exceedingly small by comparison), this mechanism will lead to the classical behavior of quantum mechanical systems for extended periods of time.

\section{Acknowledgments}
I would like to thank Ian Bouche (Boston University) for his mentorship and guidance throughout this project.  
I also thank Professor David J. Bishop, Professor Anatoli Polkovnikov, and Boston University for their support.  
Finally, I gratefully acknowledge Professor Mohamed Laradji (University of Memphis) for his mentorship and recommendation that enabled my participation in the RISE program.

\newpage

\section{Appendix}
Full derivation for the wavelike regime: 
The RHS of the differential equation can be expanded as:
\begin{align*}
\frac{\mathrm{d}W}{\mathrm{d}t} 
&= \{ H, W \}_{\mathrm{M}} \\
&= \frac{2}{\hbar} H 
   \sin\!\left( \frac{\hbar}{2} 
   \left( \overleftarrow{\partial}_x \overrightarrow{\partial}_p 
   - \overleftarrow{\partial}_p \overrightarrow{\partial}_x \right) \right) 
   W
\end{align*}

Expressing the sine operator as a Taylor series:
\begin{align*}
\frac{\mathrm{d}W}{\mathrm{d}t} = &\ H \Big[ \left( \overleftarrow{\partial}_x \!\overrightarrow{\partial}_p 
- \overleftarrow{\partial}_p \!\overrightarrow{\partial}_x \right) \\ & \quad \quad-\frac{1}{6} \left( \frac{\hbar}{2} \right)^2 
\left( \overleftarrow{\partial}_x \!\overrightarrow{\partial}_p  - \overleftarrow{\partial}_p \!\overrightarrow{\partial}_x \right)^3 
\Big] W \\
& + \mathcal{O}(\hbar^4)
\end{align*}

Applying the differential operators yields:
\begin{align*}
\frac{\mathrm{d}W}{\mathrm{d}t} = &\
 \frac{\partial H}{\partial x} \frac{\partial W}{\partial p} - \frac{\partial H}{\partial p} \frac{\partial W}{\partial x} \\
& - \frac{\hbar^2}{24} \Big[
\frac{\partial^3 H}{\partial x^3} \frac{\partial^3 W}{\partial p^3} - 3 \frac{\partial^3 H}{\partial x^2 \partial p}
 +\frac{\partial^3 W}{\partial x \partial p^2}\\&\quad\quad\quad + 3 \frac{\partial^3 H}{\partial x \partial p^2} 
\frac{\partial^3 W}{\partial x^2 \partial p} - \frac{\partial^3 H}{\partial p^3} 
\frac{\partial^3 W}{\partial x^3} 
\Big] \\
&+ \mathcal{O}(\hbar^4)
\end{align*}

The last term $\mathcal{O}(\hbar^4)$ approaches $0$ as $\hbar \to 0$, giving us:
\begin{align*}
\frac{\mathrm{d}W}{\mathrm{d}t} = &\
 \frac{\partial H}{\partial x} \frac{\partial W}{\partial p} - \frac{\partial H}{\partial p} \frac{\partial W}{\partial x} \\
& - \frac{\hbar^2}{24} \Big[
\frac{\partial^3 H}{\partial x^3} \frac{\partial^3 W}{\partial p^3} - 3 \frac{\partial^3 H}{\partial x^2 \partial p}
 +\frac{\partial^3 W}{\partial x \partial p^2}\\&\quad\quad\quad + 3 \frac{\partial^3 H}{\partial x \partial p^2} 
\frac{\partial^3 W}{\partial x^2 \partial p} - \frac{\partial^3 H}{\partial p^3} 
\frac{\partial^3 W}{\partial x^3} 
\Big]
\end{align*}

Notice that the first two terms of the RHS is just the probability rate of change for a classical particle. Hence, when the third term of the RHS is extremely small relative to the first two, the quantum particle achieves near-classical motion:
\begin{align*}
\frac{\partial H}{\partial x} \frac{\partial W}{\partial p}
- &\frac{\partial H}{\partial p} \frac{\partial W}{\partial x} \gg \\
&\ \frac{\hbar^2}{24} \Big[
\frac{\partial^3 H}{\partial x^3} \frac{\partial^3 W}{\partial p^3} 
- 3 \frac{\partial^3 H}{\partial x^2 \partial p} 
\frac{\partial^3 W}{\partial x \partial p^2} \\
&\ \quad\quad + 3 \frac{\partial^3 H}{\partial x \partial p^2} 
\frac{\partial^3 W}{\partial x^2 \partial p} \frac{\partial^3 H}{\partial p^3} 
\frac{\partial^3 W}{\partial x^3} 
\Big]
\end{align*}

Recall that 
\begin{equation*}
H = \frac{p^2}{2m} + V(x),
\end{equation*}
where $H$ represents the total energy of the system, $\frac{p^2}{2m}$ is the kinetic energy, and $V(x)$ is the potential energy as a function of position. 

The first, second, and third partial derivatives of $H$ with respect to $p$ and $x$ are:
\begin{align*}
& \frac{\partial H}{\partial p} = \frac{p}{m}, 
\frac{\partial^2 H}{\partial p^2} = \frac{1}{m}, 
\frac{\partial^3 H}{\partial p^3} = 0,
\\
& \quad \frac{\partial H}{\partial x} = V'(x),
\frac{\partial^2 H}{\partial x^2} = V''(x),
\frac{\partial^3 H}{\partial x^3} = V'''(x),
\\
& \frac{\partial^3 H}{\partial x^2 \partial p} = 0,
\quad \! \frac{\partial^3 H}{\partial x \partial p^2} = 0.
\end{align*}

We define displacements from the center of the coherent state as
\[
\delta x \equiv x - x_0, \quad \delta p \equiv p - p_0.
\]
In terms of these, the gaussian Wigner function takes the compact form
\[
W(x, p, t) = \frac{1}{\pi \hbar} 
\exp\!\left[ -\frac{\delta x^2}{\sigma_x^2} - \frac{\delta p^2}{\sigma_p^2} \right],
\]
where $\sigma_x = \sqrt{\frac{\hbar}{2 m \omega}}$ and $\sigma_p = \sqrt{\frac{\hbar m \omega}{2}}$ denote the position and momentum widths of the Gaussian state. Using this shorthand, the derivatives of \(W\) with respect to \(x\) are
\begin{align*}
\frac{\partial W}{\partial x} &= -\frac{2 \, \delta x}{\sigma_x^2} W, \\
\frac{\partial^2 W}{\partial x^2} &= \frac{4 \, \delta x^2 - 2 \sigma_x^2}{\sigma_x^4} W, \\
\frac{\partial^3 W}{\partial x^3} &= \frac{12 \sigma_x^2 \, \delta x - 8 \, \delta x^3}{\sigma_x^6} W.
\end{align*}
and similarly, the derivatives with respect to \(p\) are
\begin{align*}
\frac{\partial W}{\partial p} &= -\frac{2 \, \delta p}{\sigma_p^2} W, \\
\frac{\partial^2 W}{\partial p^2} &= \frac{4 \, \delta p^2 - 2 \sigma_p^2}{\sigma_p^4} W, \\
\frac{\partial^3 W}{\partial p^3} &= \frac{12 \sigma_p^2 \, \delta p - 8 \, \delta p^3}{\sigma_p^6} W,
\end{align*}

Substituting these differentials into the inequality yields:
\begin{align*}
&\frac{\hbar^2}{24} V'''(x) 
\frac{12 \sigma_p^2 \delta p - 8 \delta p^3}{\sigma_p^6} \ll \frac{2 p}{m} \frac{\delta x}{\sigma_x^2} 
- 2 V'(x) \frac{\delta p}{\sigma_p^2}
\end{align*}

Which in terms of $m$, $\omega$ and $\hbar$ yields: 
\begin{align*}
&\frac{(3\hbar m \omega\delta p - 4(\delta p)^3) V'''(x)}
{6m^2\omega^2(m\omega^2p\delta x - \delta pV'(x))} \ll 1
\end{align*}

\end{document}